\documentclass[pre,twocolumn,showpacs]{revtex4}
\usepackage{epsfig}
\usepackage{amssymb}
\begin{document}

\title{
Two different kinds of rogue waves in weakly-crossing  sea states} 
\author{V.~P. Ruban}
\affiliation{Landau Institute for Theoretical Physics,
2 Kosygin Street, 119334 Moscow, Russia} 

\date{\today}

\begin{abstract}
Formation of giant waves in sea states with two spectral maxima, centered at 
close wave vectors ${\bf k}_0\pm\Delta {\bf k}/2$ in the Fourier plane,
is numerically simulated using the fully nonlinear model for long-crested 
water waves [V. P. Ruban, Phys. Rev. E {\bf 71}, 055303(R) (2005)].
Depending on an angle $\theta$ between  the vectors ${\bf k}_0$ and 
$\Delta {\bf k}$, which determines a typical orientation of interference 
stripes in the physical plane, rogue waves arise having different 
spatial structure. If $\theta \lesssim\arctan(1/\sqrt {2})$, then 
typical giant waves are relatively long fragments of 
essentially two-dimensional (2D) ridges, separated by wide valleys and 
consisting of alternating oblique crests and troughs. 
At nearly perpendicular ${\bf k}_0$ and $\Delta {\bf k}$, 
the interference minima develop to coherent structures
similar to the dark solitons of the nonlinear Shroedinger equation, 
and a 2D freak wave looks much as a piece of a 1D freak wave, 
bounded in the transversal direction by two such dark solitons.

\end{abstract}




\maketitle


The problem of extreme ocean waves (known as freak waves, rogue waves, 
or killer waves) has attracted much attention in recent years 
(see, e.g., the reviews \cite{Kharif-Pelinovsky, DKM2008}, 
where different physical mechanisms of the rogue wave phenomenon are discussed, 
and many related works are referenced; for some recent developments in this 
field, see Refs.~\cite{RogueWaves2006,Janssen2003,S-J_D_T_K_L-2005, 
OOS2006,SKEMS2006,GT2007,FGD2007,R2006PRE,R2007PRL,ZDV2002,
DZ2005Pisma,ZDP2006,DZ2008,FT2009,Onorato_et_al_2009}).
With a typical background wave amplitude 
$A_0\approx [0.015\cdots 0.02]\lambda_0$
(where $\lambda_0=2\pi/k_0$ is a typical wave length),
the maximum elevation of a freak wave can reach a height
$Y_{\scriptsize\mbox{max}}>0.06\lambda_0$, which approaches the limiting 
Stokes wave. Profiles of freak waves are very steep, and they strongly deviate 
from the sinusoidal shape. In different circumstances, the giant waves 
can be caused by different reasons. Accordingly, there are several probable
scenarios explaining formation of these waves. It has been recognized that 
one of the most important reasons for freak waves is the modulational 
Benjamin-Feir-Zakharov instability taking place in relatively long and high 
groups of propagating waves \cite{Zakharov67,Benjamin-Feir}. 
Efficiency of this mechanism is usually characterized by the so-called 
Benjamin-Feir index (BFI) \cite{Janssen2003},
\begin{equation}
\mbox{BFI}\sim  \lambda_0^{-2} A_0 l_0,
\end{equation}
where $l_0$ is a typical length of wave groups.
For example, in completely incoherent sea states (low BFI) the
modulational instability is suppressed, and rather rare
appearance of anomalous wave events is basically of a purely kinematic origin.
This limit is well described by the approximation of non-interacting normal
wave modes, renormalized by a weakly-nonlinear transformation
excluding three-wave (non-resonant) processes (see, e.g., 
Refs.\cite{T1980,DKM2008,FT2009}, and references therein).
Higher values of BFI correspond to more favorable conditions for the occurrence 
of freak waves. Nonlinear wave interactions become essential, 
so giant waves arise in the process of evolution 
of some coherent structures. In particular, the limit of 
infinitely high BFI (a weakly disturbed planar wave as an initial state)
has been recently studied in works \cite{R2006PRE,R2007PRL}, 
and specific zigzag-shaped, obliquely oriented wave stripes were found 
to develop in the nonlinear stage of the modulational instability, 
with rogue waves occurring mainly at zigzag turns. 
That case roughly corresponds to another
probable scenario, when refraction of swell in a spatially non-uniform 
current causes significant preliminary amplification of wave height 
around caustic region \cite{Peregrine,LP2006}. 
A different kind of coherent structures has been recognized recently for 
purely one-dimensional (1D) waves (planar flows), 
the so-called giant breathers \cite{DZ2008}, 
which are extremely short and steep envelope solitons, containing just 
one-two waves.

Thus, though BFI is definitely a relevant parameter, but in some situations 
it does not completely characterize the freak waves, as it takes place, 
for example, for 1D waves \cite{ZDV2002,DZ2005Pisma,ZDP2006,DZ2008},
or for very long-crested waves \cite{GT2007,Onorato_et_al_2009}, 
or in crossing sea states \cite{OOS2006,SKEMS2006}. The reason is that 
coherent wave structures depend on additional parameters as well.
In the present work, we investigate this question in more detail
for weakly crossing sea states. More specifically, we consider sea states
with two spectral peaks centered at wave vectors ${\bf k}_0-\Delta {\bf k}/2$ 
and ${\bf k}_0+\Delta {\bf k}/2$ in the Fourier plane, and we assume  
$|\Delta {\bf k}|\ll|{\bf k}_0|$. 
Such a situation corresponds to the presence of relatively long and 
wide wave stripes obliquely oriented to the wave fronts in a range of angles 
near the angle $\theta$ between the vectors $\Delta {\bf k}$ and ${\bf k}_0$.
We study numerically how the process of rogue wave formation depends 
on the angle $\theta$. The computations are based on the approximate 
theoretical model for long-crested fully nonlinear water waves, developed in 
Refs.\cite{R2005PRE,RD2005PRE} and later successfully applied 
in Refs.\cite{R2006PRE,R2007PRL}. The model is intermediate 
between the exact Eulerian dynamics and the approximate equations for wave 
envelopes [generalizations of the nonlinear Shroedinger equation (NLSE)] 
suggested in Refs.\cite{Dysthe1979,TD1996,TKDV2000}. 
It should be emphasized that our method makes possible to compute 
profiles of individual waves, while in the works \cite{OOS2006,SKEMS2006}
only wave envelopes in crossing sea states for $\theta=\pi/2$ were studied.

\begin{figure}
\begin{center}
\epsfig{file=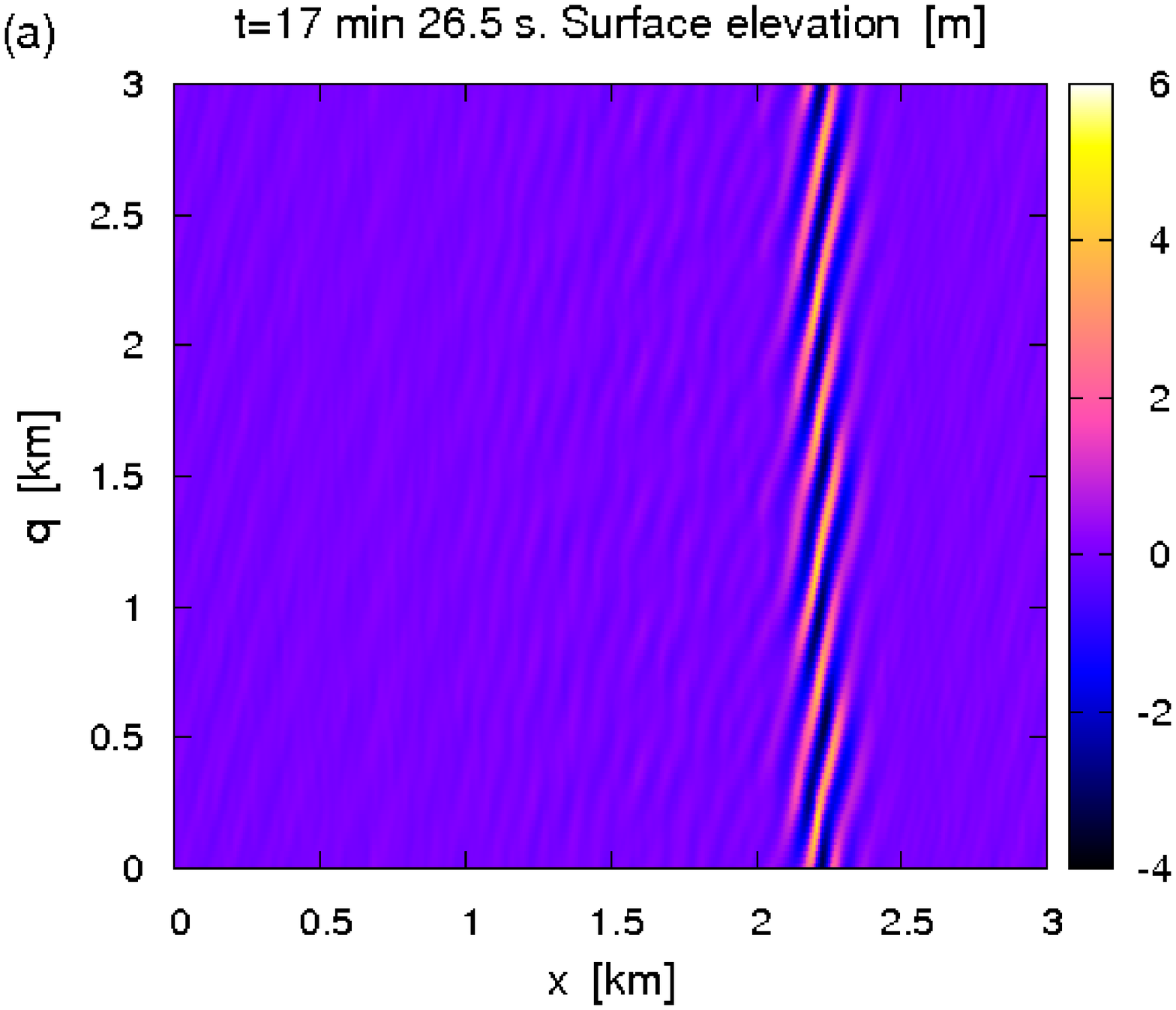,width=70mm}\\
\epsfig{file=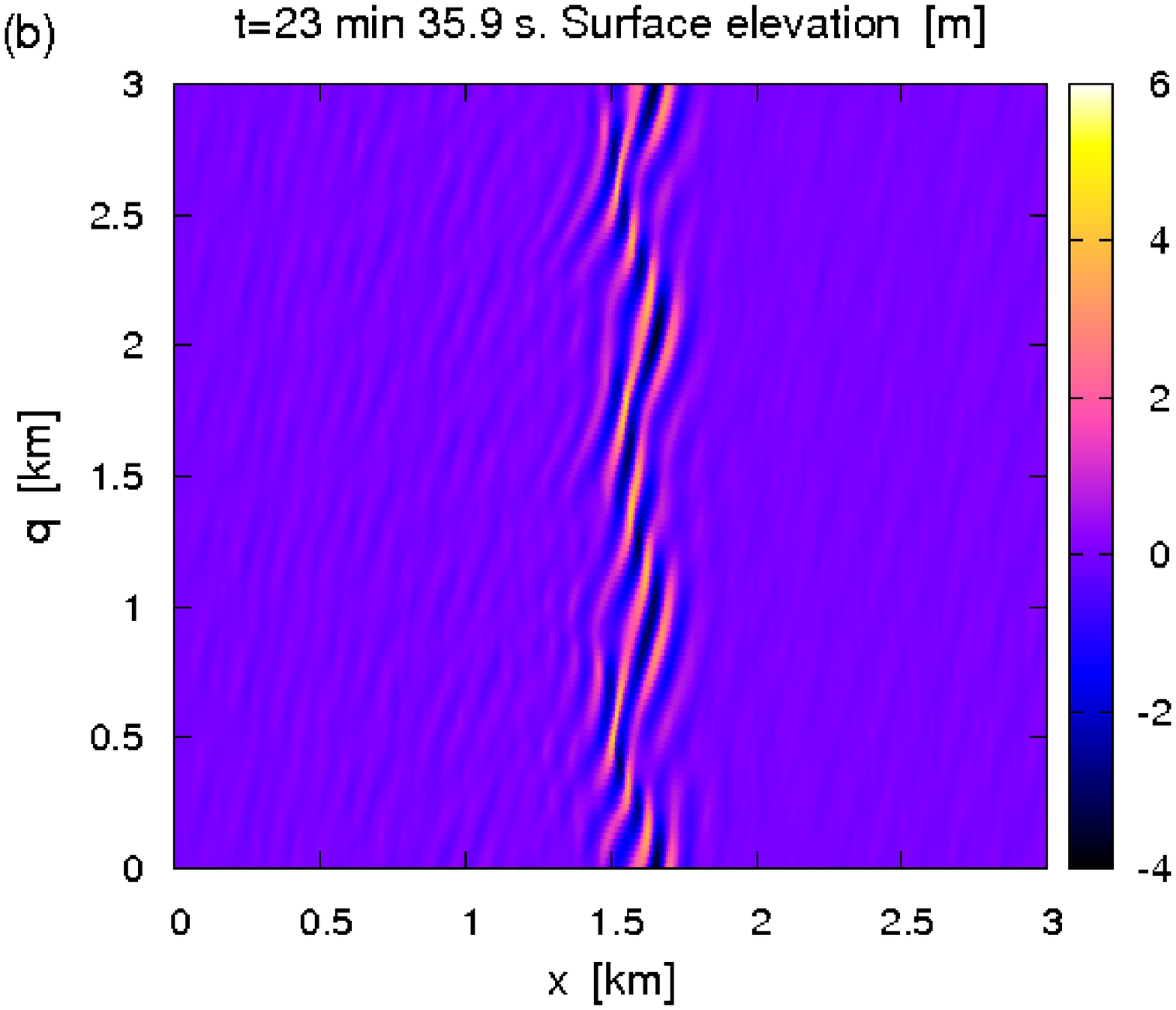,width=70mm}
\end{center}
\caption{(Color online). Evolution of a perturbed high-amplitude oblique 
soliton into a zigzag structure.} 
\label{R_16_t1_t2} 
\end{figure}

\begin{figure}
\begin{center}
   \epsfig{file=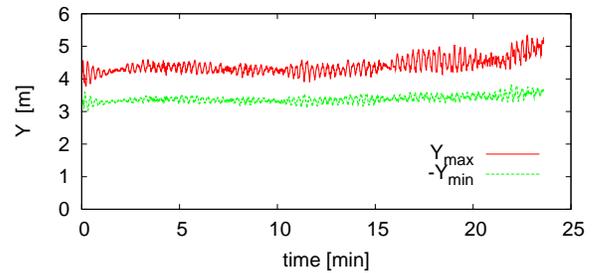,width=80mm}  
\end{center}
\caption{(Color online). Maximum and minimum elevation of 
the oblique soliton vs. time.} 
\label{R_16Ymax} 
\end{figure}

The main results of this work are the following. 
If $\theta \lesssim\arctan(1/\sqrt {2})$, then the nonlinearity is 
defocusing along the stripes and it is focusing across them.
The situation is opposite at nearly perpendicular ${\bf k}_0$ and 
$\Delta {\bf k}$. Accordingly, rogue waves, occurring at the sea surface, 
are different in these two cases. In the first case, freak waves look 
as fragments of structures which are similar to the solitonic 
solutions  of a focusing NLSE for a wave envelope.
Such extremely narrow and steep solitons are in essence rows 
of alternating oblique crests and troughs (see Fig.\ref{R_16_t1_t2}a).
When perturbed by a weak two-dimensional (2D) random field, 
the oblique solitons can exist for many wave periods almost unchanged, 
but later they transform to zigzag structures similar to that 
described in Ref.\cite{R2007PRL}
(see Figs.\ref{R_16_t1_t2}-\ref{R_16Ymax} for example).
It should be noted that in the limit $\theta\to 0$ such extreme oblique 
solitons coincide with the recently discovered 1D giant breathers 
\cite{DZ2008}. Thus, the fundamental role of these coherent structures 
in the dynamics of water waves is confirmed.
At nearly perpendicular ${\bf k}_0$ and $\Delta {\bf k}$,
another kind of coherent structures comes into play, similar to the
dark solitons of a defocusing NLSE. Dark solitons develop at the 
interference minima and they transversally separate wave groups 
subjected to the longitudinal modulational instability.
Freak waves in this case have nearly 1D profiles, but they are
bounded in the transversal direction by two dark solitons.


To understand better numerical results, it is useful to have in mind 
a qualitative model describing weakly nonlinear water waves in terms 
of a complex wave amplitude $A(x_1, x_2,t)$, which determines the 
free surface elevation as follows,
\begin{equation}
Y(x_1, x_2,t)\approx \mbox{Re}\left[
A(x_1, x_2,t)\exp(ik_0 x_1-i\omega_0 t)\right],
\end{equation}
where $x_1$ and $x_2$ are horizontal coordinates, with $x_1$ along ${\bf k}_0$,
$\omega_0=(g k_0)^{1/2}$ is frequency of the carrier wave,
and $g$ is the gravity acceleration.
The function $A(x_1, x_2,t)$ is known to approximately obey a 2D NLSE 
\cite{Zakharov67}, 
\begin{equation}\label{NLSE_2D}
\frac{i}{\omega_0}\frac{\partial A}{\partial t}
+\frac{i}{2k_0}\frac{\partial A}{\partial x_1}=
\frac{1}{8k_0^2}\left(\frac{\partial^2 A}{\partial x_1^2}
-2\frac{\partial^2 A}{\partial x_2^2}\right)+\frac{k_0^2}{2}|A|^2A.
\end{equation}
The oblique stripes  roughly correspond to the following 
1D reductions of Eq.(\ref{NLSE_2D}):
\begin{equation}\label{NLSE_1D_reductions}
A=k_0^{-1}\Psi(\xi, \tau),
\quad \xi= k_0[(x_1-V_{\scriptsize\mbox{gr}}t)\cos\theta + x_2\sin\theta],
\end{equation}
where $\tau=\omega_0 t$, and $V_{\scriptsize\mbox{gr}}=(\omega_0/2k_0)$ 
is the group velocity.
As a result, we have a 1D NLSE describing the transversal dynamics of 
idealized, infinitely long wave stripes,
\begin{equation}\label{NLSE_1D}
i\Psi_\tau=
\frac{1}{4}\left[(1/2)\cos^2\theta-\sin^2\theta\right]
\Psi_{\xi\xi}+\frac{1}{2}|\Psi|^2\Psi.
\end{equation}
Depending on the sign of the dispersion coefficient 
$\alpha(\theta)=[(1/2)\cos^2\theta-\sin^2\theta]$, this is either focusing
equation or defocusing one, and the dynamics is quite different in each case. 
For example, in the focusing case (when $\alpha>0$), 
the nonlinearity can become saturated with the so-called (bright) solitons,
\begin{equation}\label{solitons}
\Psi_{\scriptsize\mbox{bs}}=
\frac{s}{\cosh\left[(s/\sqrt\alpha)(\xi-\xi_0)\right]}
\exp(-i \tau s^2/4+i\phi_0),
\end{equation}
where $s$ is a wave steepness, and $\xi_0$, $\phi_0$ are arbitrary constants.
These solutions describe  infinitely long wave ridges 
consisting of alternating oblique crests and troughs. 
Physical conditions of applicability of the above formula imply 
$s\lesssim 0.1$ and $s/\sqrt{\alpha}\ll 1$, but actually these solutions 
have been found to continue qualitatively to considerably higher values 
$s\lessapprox 0.27$.
We specially studied  a long-time behaviour of such extreme solitons,
both for $\theta=0$ (the giant breathers at 2D surface), 
and for $\theta\not=0$. 
It is one of the main results of the present work that 
in two dimensions extreme solitons can exist for a 
long time before transformation into zigzag structures. 
An example of evolution of a perturbed high-amplitude oblique soliton 
is presented in Figs.\ref{R_16_t1_t2}-\ref{R_16Ymax}, for 
$\lambda_0\approx 100$ m, $\theta\approx \arctan(1/5)$, and $s\approx 0.22$.

\begin{figure}
\begin{center}
   \epsfig{file=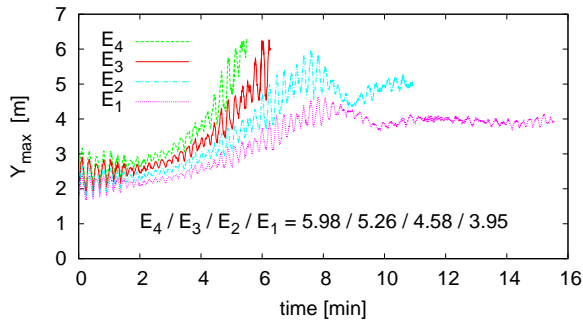,width=80mm}  
\end{center}
\caption{(Color online). Maximum elevation of the free surface vs. time
in the numerical experiments A1-A4.} 
\label{CS_20Ymax} 
\end{figure}
\begin{figure}
\begin{center}
\epsfig{file=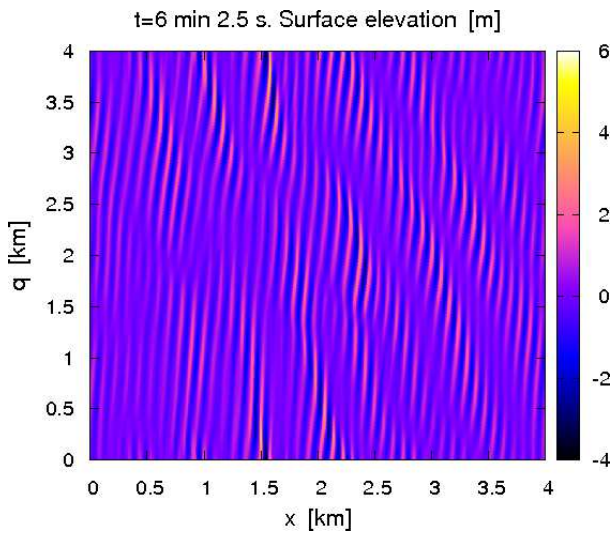,width=80mm}    
\end{center}
\caption{(Color online). Experiment A3: 
the two big waves are at $x\approx 1.6$ km, $q\approx [3.7\cdots 3.9]$ km, 
and at $x\approx 1.5$ km, $q\approx [0.1\cdots 0.3]$ km.} 
\label{CS_20map} 
\end{figure}

In the defocusing case (when $\alpha<0$), the so-called dark solitons
are possible,
\begin{equation}\label{dark_solitons}
\Psi_{\scriptsize\mbox{ds}}=
s\tanh\left[(s/\sqrt{-\alpha}) (\xi-\xi_0)\right]
\exp(-i \tau s^2/2+i\phi_0),
\end{equation}
which separate two domains of opposite amplitude.

In view of the above, it is clear that since the effective dispersion  
coefficient  $\alpha(\theta)$ changes the sign at 
$\theta_*=\arctan(1/\sqrt 2)$,
in the full 2D dynamics of random wave fields there  should be two 
substantially different regimes, one regime at $\theta\lesssim\theta_*$ 
and another at $\theta$ close to $\pi/2$. 
This hypothesis is confirmed in general by numerical experiments reported here.

\begin{figure}
\begin{center}
   \epsfig{file=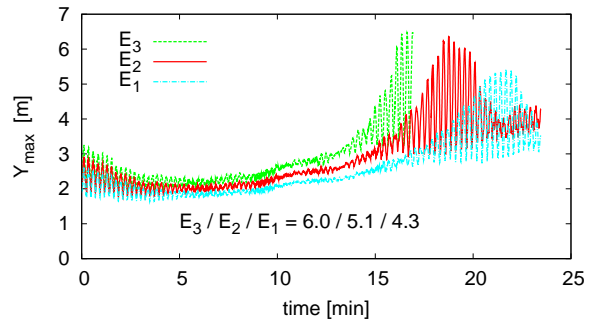,width=80mm}  
\end{center}
\caption{(Color online). Maximum elevation of the free surface
in the numerical experiments B1-B3.} 
\label{CS_40Ymax} 
\end{figure}
\begin{figure}
\begin{center}
\epsfig{file=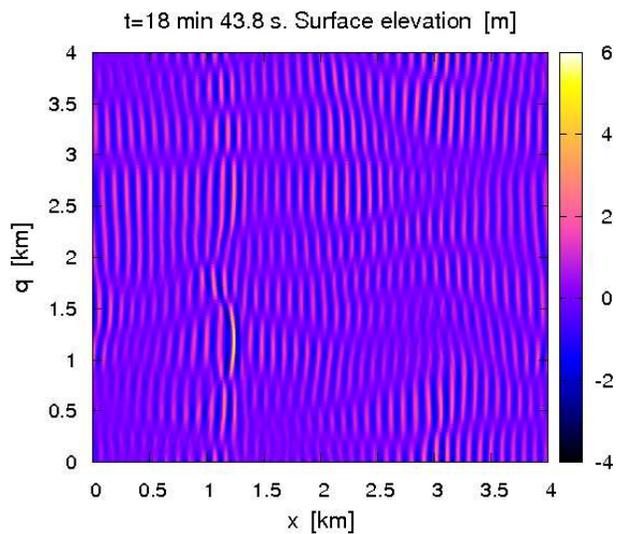,width=80mm}
\end{center}
\caption{(Color online). Experiment B2: 
the rogue wave is at $x\approx 1.2$ km, $q\approx [1.0\cdots 1.4]$ km.} 
\label{CS_40map} 
\end{figure}
\begin{figure}
\begin{center}
   \epsfig{file=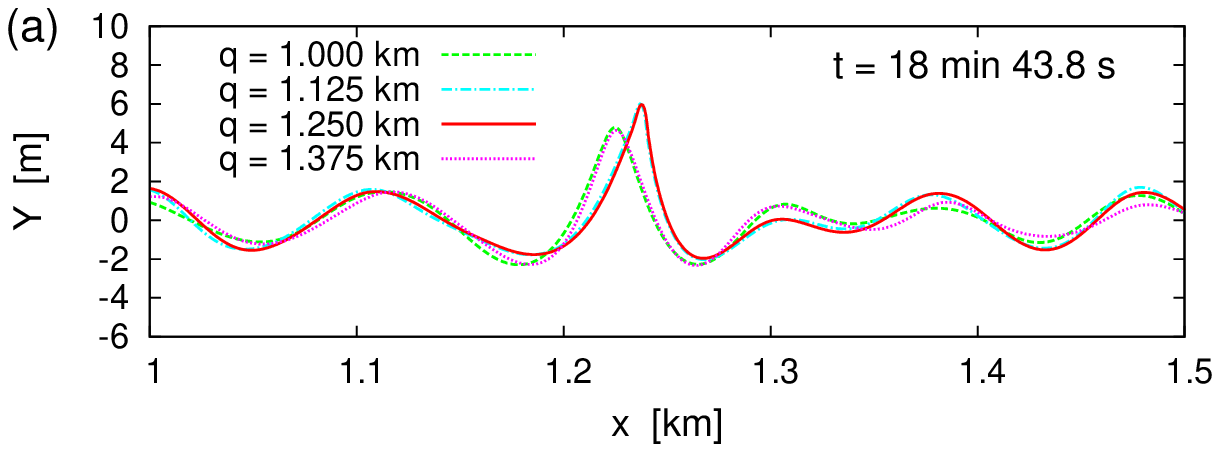,width=83mm}\\  
   \epsfig{file=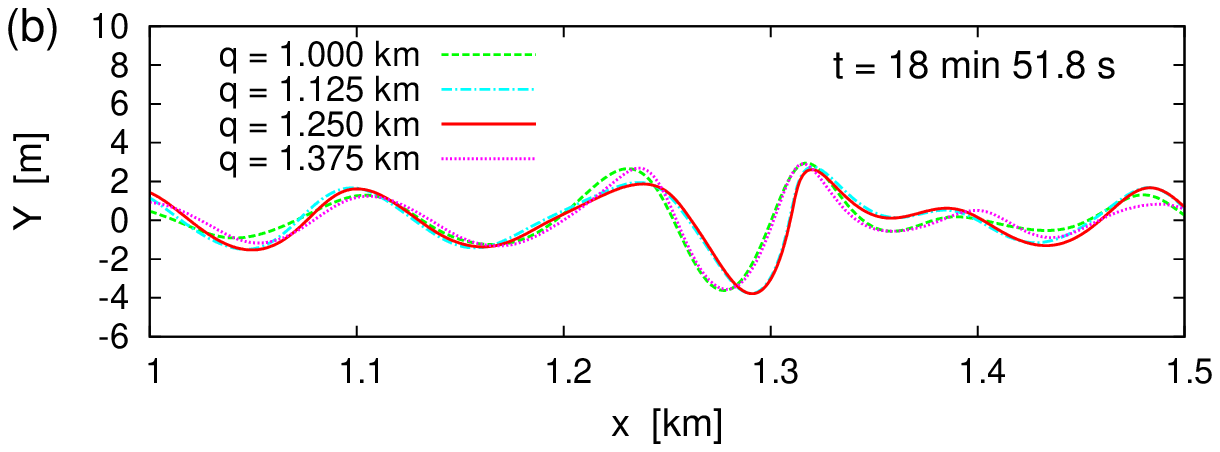,width=83mm}\\
   \epsfig{file=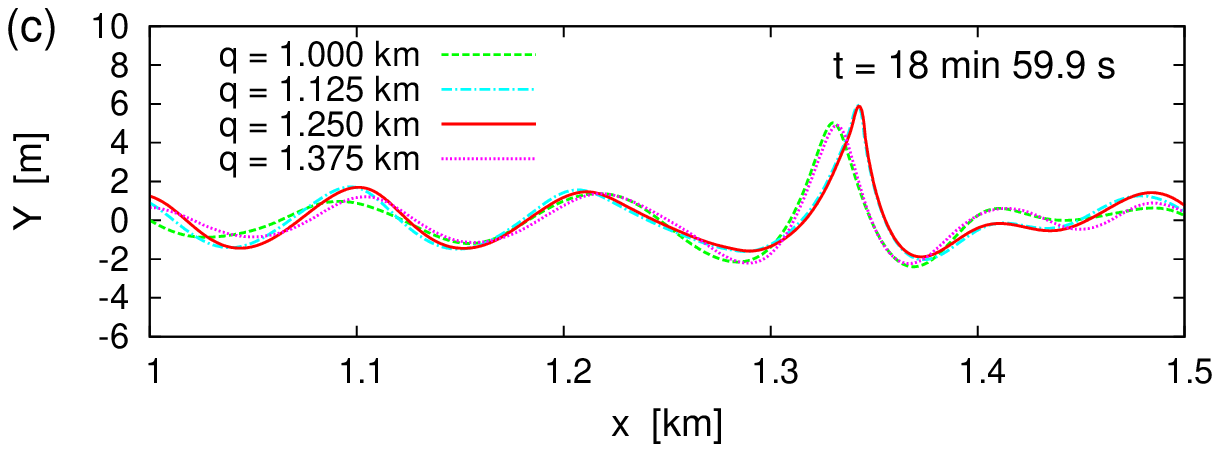,width=83mm} 
\end{center}
\caption{(Color online). (a) profiles of the freak wave from 
Fig.\ref{CS_40map}; (b) 8 s later: ``a hole in the sea'';
(c) 16 s later: the big wave has risen again.} 
\label{CS_40_t1_t2_t3} 
\end{figure}

The computations were performed in the dimensionless square domain 
$2\pi\times 2\pi$ with periodic boundary conditions along the horizontal 
coordinates $x$ and $q$ (see  Refs.\cite{R2005PRE,RD2005PRE} for details).
Thus, all the discrete Fourier modes correspond to integer wave vectors 
${\bf k}=(k,m)$. 
The vector ${\bf k}_0$ was generally taken slightly different from the 
direction of $x$ axis, in order to take into account the effect of gradual 
re-orientation of wave crests along the oblique stripes 
(see Ref.\cite{R2007PRL}). Final results were 
rescaled to give a convenient for presentation value $\lambda_0\approx 100$ m, 
which is quite typical in natural sea conditions. The corresponding wave period
is $T_0=[2\pi\lambda_0/g]^{1/2}\approx 8$ s. Two small sets of typical 
numerical experiments are presented, designated as A1-A4 and B1-B3. 
Within each set, at $t=0$ the normal Fourier modes of the wave 
field were taken in the form $a_{km}(0)=c F(k,m)\exp(i\gamma_{km})$, 
with a positive function $F(k,m)$ having two nearly Gaussian maxima at 
${\bf k}_0\pm\Delta {\bf k}/2$, and with quasi-random initial phases 
$\gamma_{km}$ (in set A and in set B the phases $\gamma_{km}$ were not 
the same). In each experiment the coefficient $c$ was different, 
thus resulting in different values of the total energy $E$.
In set A we chose ${\bf k}_0=(40.0, -2.5)$ and $\Delta {\bf k}=(7.0, 2.0)$, 
so a case $\theta<\theta_*$ was simulated, while
in set B it was a crossing sea state with $\theta=\pi/2$:
${\bf k}_0\pm\Delta {\bf k}/2=(39.5, \pm 3.5)$.

For set A, some results are presented in Figs.\ref{CS_20Ymax}-\ref{CS_20map}.
The modulational instability acts in this case from the very beginning,
and it needs a short time $5..8$ min to produce freak waves in the initially 
most tall wave groups. 
The two neighbouring big waves in Fig.\ref{CS_20map} look as a fragment 
of an oblique soliton (compare to Fig.\ref{R_16_t1_t2}a).
The computations A4 and A3 were terminated at the moments when the freak waves 
broke, while in experiments A2 and A1 the waves remained smooth, so at later
times nearly stationary, long oblique solitons were observed (not shown).
 
Results of experiments B1-B3 (see  Figs.\ref{CS_40Ymax}-\ref{CS_40_t1_t2_t3})
are more intriguing, since there were two stages
in the evolution of the wave field before rogue waves arose. In the first 
stage, for $4..7$ min after the beginning, it was the formation of dark 
solitons along interference minima, which process was accompanied by 
substantial decreasing of the wave amplitude along interference maxima. 
At the end of this stage, the free surface has been divided by dark solitons 
into domains of nearly 1D dynamics.
In the second stage, adjacent domains interact in a complicated manner, 
and in one of them the amplitude increases, resulting in fast development 
of the longitudinal modulational instability. As the result, 
a single rogue wave grows, which is squeezed from the lateral sides 
between two dark solitons, as shown in Fig.\ref{CS_40map}. 
The rogue wave is ``breathing'', with time-alternating tall crest and deep 
trough, and it approximately repeats the profile after $2T_0$ 
(see Fig.\ref{CS_40_t1_t2_t3}).
After a dozen of the oscillations, the big wave spreads in the transversal
direction and disappears (not shown). 

As results A and B are compared, it becomes clear that the unusual properties 
of the abnormal waves in weakly-crossing seas are more prominent when 
the interference stripes are nearly perpendicular to the wave crests.

{\it Acknowledgments.}
These investigations were supported by RFBR (grants 09-01-00631 
and 07-01-92165),
by the ``Leading Scientific Schools of Russia'' grant 4887.2008.2,
and by the Program ``Fundamental Problems of Nonlinear Dynamics'' 
from the RAS Presidium.

\end{document}